\documentclass{article}
\usepackage{spconf,amsmath,graphicx,amssymb}
\usepackage{amsfonts}
\usepackage{bbm}
\usepackage{booktabs}
\usepackage{hyperref}

\renewenvironment{thebibliography}[1]{
  \begin{oldthebibliography}{#1}
    \vspace{-10pt}
    \setlength{\itemsep}{.35em}
    \setlength{\parskip}{1pt}
}
{
  \end{oldthebibliography}
}

\title{Learning contextual tag embeddings for cross-modal alignment of audio and tags}
\name{Xavier Favory$^\dagger$, Konstantinos Drossos$^\ddagger$, Tuomas Virtanen$^\ddagger$, Xavier Serra$^\dagger$\thanks{The authors would like to thank all the Freesound users that have been sharing very valuable content for many years.
X. Favory and K. Drossos are grateful for the GPUs donated by NVidia. K. Drossos and T. Virtanen wish to acknowledge CSC-IT Center for Science, Finland, for computational resources.}}
\address{$^\dagger$Music Technology Group, Universitat Pompeu Fabra, Spain\\
$^\ddagger$Audio Research Group, Tampere University, Finland}
\begin{document}
%
\maketitle
\begin{abstract}
Self-supervised audio representation learning offers an attractive alternative for obtaining generic audio embeddings, capable to be employed into various downstream tasks. Published approaches that consider both audio and words/tags associated with audio do not employ text processing models that are capable to generalize to tags unknown during training. In this work we propose a method for learning audio representations using an audio autoencoder (AAE), a general word embeddings model (WEM), and a multi-head self-attention (MHA) mechanism. MHA attends on the output of the WEM, providing a contextualized representation of the tags associated with the audio, and we align the output of MHA with the output of the encoder of AAE using a contrastive loss. We jointly optimize AAE and MHA and we evaluate the audio representations (i.e. the output of the encoder of AAE) by utilizing them in three different downstream tasks, namely sound, music genre, and music instrument classification. Our results show that employing multi-head self-attention with multiple heads in the tag-based network can induce better learned audio representations.

\end{abstract}
\begin{keywords}
representation learning, multimodal contrastive learning, audio classification
\end{keywords}

\vspace{-6pt}
\section{Introduction}
\vspace{-6pt}
\label{sec:intro}
An effective way to learn audio representations that can be used for sound classification involves training deep neural networks (DNNs) on supervised tasks, using large annotated datasets~\cite{gemmeke2017audio, lee2018samplecnn, pons2019musicnn}.
However, these datasets require a considerable amount of effort to be built
and are always limited in size, hindering the performance of learned representations.
Recent research approaches explore and adopt unsupervised, self-supervised or semi-supervised learning methods for obtaining generic audio representations, that later can be used for different downstream tasks%
~\cite{cramer2019look, jansen2018unsupervised, alwassel2019self, turpault2019semi}.
The large amount of multimedia data available online is a great opportunity for these types of approaches to learn powerful audio representations.

In the natural language and image processing fields, both supervised and unsupervised approaches enabled the creation of powerful pre-trained models, that are often employed in many different tasks~\cite{mikolov2013efficient, devlin2018bert, chen2020simple}.
%
%
Contrastive learning recently got a lot of attention due to its success for the unsupervised pre-training of DNNs, enabling to learn flexible representation without the need of having labels associated with the content~\cite{chen2020simple,le2020contrastive}.
Only the definition of positive and negative data pairs is required in order to learn a model that will produce a latent space reflecting semantic characteristics.
%
The association of images and text can be exploited for learning embeddings~\cite{wu2019learning}, e.g. by 
using a self-attention mechanism to learn context sensitive text embeddings that are then aggregated into sentence embeddings.
%
%
Similar approaches have been adopted for machine listening, for instance unsupervised pre-training of transformer models can improve speech recognition performance by employing contrastive predictive coding strategies~\cite{jiang2019improving,oord2018representation}.
In \cite{turpault2019semi}, the authors employ a semi-supervised sampling strategy to create triplets for benefiting automatic tagging systems. However, these approaches consider just one modality, i.e. audio. 

A cross-modal method for learning and aligning audio and tag latent representations (COALA) was presented in~\cite{favory2020coala}.
Latent representations were learnt using autoencoders, one aiming at encoding and reconstructing spectrogram representations of sounds, and the other focusing on encoding and reconstructing a set of tags represented as multi-hot vectors.
The outcomes suggest that it is possible to leverage noisy, user-generated data of audio and accompanying tags, for learning semantically enriched audio representations that can be used for different classification tasks.
%
However, in COALA the tag-based input representation was fixed, and therefore the tag-based encoder cannot generalize to now terms that have not been seen during training.
This makes the approach loose the flexibility of contrastive representation learning. 
%

In this work we propose a method for allowing the textual generalization of cross-modal approaches, 
using pre-trained word embedding models which project words into semantic spaces. 
We propose an attention mechanism for learning higher-level contextualized semantic representation similarly to~\cite{wu2019learning}.
However, our approach relies on accompanying tags instead of text, and therefore employs a simpler approach for computing semantic embeddings.
The rest of the paper is as follows. 
In Section 2 we present our proposed method. 
Section 3 describes the utilized dataset, the tasks and metrics that we employed for the assessment of the performance, the baselines that we compare our method with.
The results of the evaluation is presented and discussed in Section 4. 
Finally, Section 5 concludes the paper and proposes future research directions.

\vspace{-6pt}
\section{Proposed Method}
\vspace{-6pt}
\label{sec:method}
Our method, illustrated in Figure 1, consists of an audio encoder, $e_{\text{a}}(\cdot)$, an audio decoder, $d_{\text{a}}(\cdot)$, a pre-trained word embedding model, $e_{\text{w}}(\cdot)$, and multi-head self-attention, $\text{Att}(\cdot)$. As input to our method, we employ a dataset of $N_{\text{B}}$ paired exampled examples, $\mathbb{G}=\{(\mathbf{X}_{\text{a}}^{n_{\text{B}}}, \mathbf{x}^{n_{\text{B}}}_{\text{w}})\}_{{n_{\text{B}}}=1}^{N_{\text{B}}}$, where $\mathbf{X}_{\text{a}}\in\mathbb{R}^{T_{\text{a}}\times F_{\text{a}}}$ is a time-frequency audio representation of $T_{\text{a}}$ audio feature vectors of $F_{\text{a}}$ features, and $\mathbf{x}_{\text{w}}=\{x^{i}_{\text{w}}\}_{i=1}^{T_{\text{w}}}$ is a set of $T_{\text{w}}$ tags, $x^i_{w}$, like ``techno'' and ``dog bark'', and associated with $\mathbf{X}_{\text{a}}$\footnote{For the clarity of notation, the superscript $n_{\text{B}}$ is dropped here and for the rest of the document, unless it is explicitly needed.}. 
Att extracts from $\mathbf{x}_{\text{w}}$ an embedding containing the context of the tags, and by using a contrastive loss we align the output $\mathbf{z}_{\text{a}} = e_{\text{a}}(\mathbf{X}_{\text{a}})$ with the output of Att. $\mathbf{z}_{\text{a}}$ is also used as an input to $\mathbf{d}_{\text{a}}$, to reconstruct $\mathbf{X}_{\text{a}}$, similarly to~\cite{favory2020coala}, effectively infusing $\mathbf{z}_{\text{a}}$ with both semantic (from $\mathbf{x}_{\text{w}}$) and low-level acoustic information (from the reconstruction by $\mathbf{d}_{\text{a}}$). 
The code of our method is available online\footnote{\url{https://github.com/xavierfav/ae-w2v-attention}}.
%
%
%
%
\vspace{-6pt}
\subsection{Audio encoding and decoding}
The encoder $e_{\text{a}}$ consists of $N_{\text{e-a}}$ cascaded 2D convolutional neural networks (CNNs), $\text{CNN}_{n_{\text{e-a}}}$, with $C^{\text{in}}_{n_{\text{e-a}}}$ and $C^{\text{out}}_{n_{\text{e-a}}}$ input and output channels, respectively, a square kernel of $K_{\text{e-a}}$ size, and $S_{\text{e-a}}$ stride. The CNNs process $\mathbf{X}_{\text{a}}$ is a serial fashion, and each $\text{CNN}_{n_{\text{e-a}}}$ is followed by a batch normalization process ($\text{BN}_{n_{\text{e-a}}}$), a rectified linear unit (ReLU), and a dropout (DO) with probability $p_{\text{a}}$, as 
\begin{equation}
    \mathbf{H}_{n_{\text{e-a}}} = \text{DO}(\text{ReLU}(\text{BN}_{n_{\text{e-a}}}(\text{CNN}_{n_{\text{e-a}}}(\mathbf{H}_{n_{\text{e-a}}-1}))))\text{,}
\end{equation}
\noindent
where $\mathbf{H}_{0}=\mathbf{X}_{\text{a}}$. $\mathbf{H}_{N_{\text{e-a}}}\in\mathbb{R}_{\geq0}^{C^{\text{e-out}}_{N_{\text{e-a}}}\times T'_{N_{\text{e-a}}}\times F'_{N_{\text{e-a}}}}$ is flattened to a vector and given as an input to a layer normalization process (LN)~\cite{ba2016layer} ($\text{FFN}_{\text{e-a}}$), as $\mathbf{z}_{\text{a}} = \text{LN}(\mathbf{h}_{N_{\text{e-a}}})\text{,}$
where $\mathbf{h}_{N_{\text{e-a}}}$ is the flattened $\mathbf{H}_{N_{\text{e-a}}}$, and $\mathbf{z}_{\text{a}}\in\mathbb{R}^{V}$, with $V = C^{\text{e-out}}_{N_{\text{e-a}}}\cdot T'_{N_{\text{e-a}}}\cdot F'_{N_{\text{e-a}}}$, is the learned audio embedding by our method. $\mathbf{z}_{\text{a}}$ is used at the employed contrastive loss, in order to be aligned with the information contained at the associated tags $\mathbf{x}_{\text{w}}$, and as an input to $d_{\text{a}}$ in order to encode low-level acoustic features in $\mathbf{z}_{\text{a}}$, similarly to~\cite{favory2020coala}.

The decoder $d_{\text{a}}$ takes as an input $\mathbf{z}_{\text{a}}$ and processes it through a series of $N_{\text{e-a}}$ transposed 2D CNNs~\cite{radford:2015:iclr,dumoulin:2016:guide}, $\text{CNN}_{n_{\text{d-a}}}$, in a reverse fashion to $e_{\text{a}}$. Firstly $\mathbf{z}_{\text{a}}$ is turned to the matrix $\mathbf{Z}_{\text{a}}\in\mathbb{R}^{C^{\text{e-out}}_{N_{\text{e-a}}}\times T'_{N_{\text{e-a}}}\times F'_{N_{\text{e-a}}}}$ and then is processed by $\text{CNN}_{n_{\text{d-a}}}$ as 
\begin{equation}
    \mathbf{H}_{n_{\text{d-a}}} = \text{ReLU}(\text{BN}_{n_{\text{e-a}}}(\text{CNN}_{n_{\text{d-a}}}(\text{DO}(\mathbf{H}_{n_{\text{d-a}}-1}))))\text{,}
\end{equation}
\noindent
where $\mathbf{H}_{0} = \mathbf{Z}_{\text{a}}$ and $\text{BN}_{n_{\text{e-a}}}$ is the batch normalization process used after $\text{CNN}_{n_{\text{d-a}}}$. The reconstructed input, $\hat{\mathbf{X}}_{\text{a}}$, is obtained as $\hat{\mathbf{X}}_{\text{a}} = \sigma(\mathbf{H}_{N_{\text{d-a}}})\text{,}$
where $\sigma$ is the sigmoid function. 
\begin{figure}
  \centering
  \includegraphics[clip,width=\columnwidth]{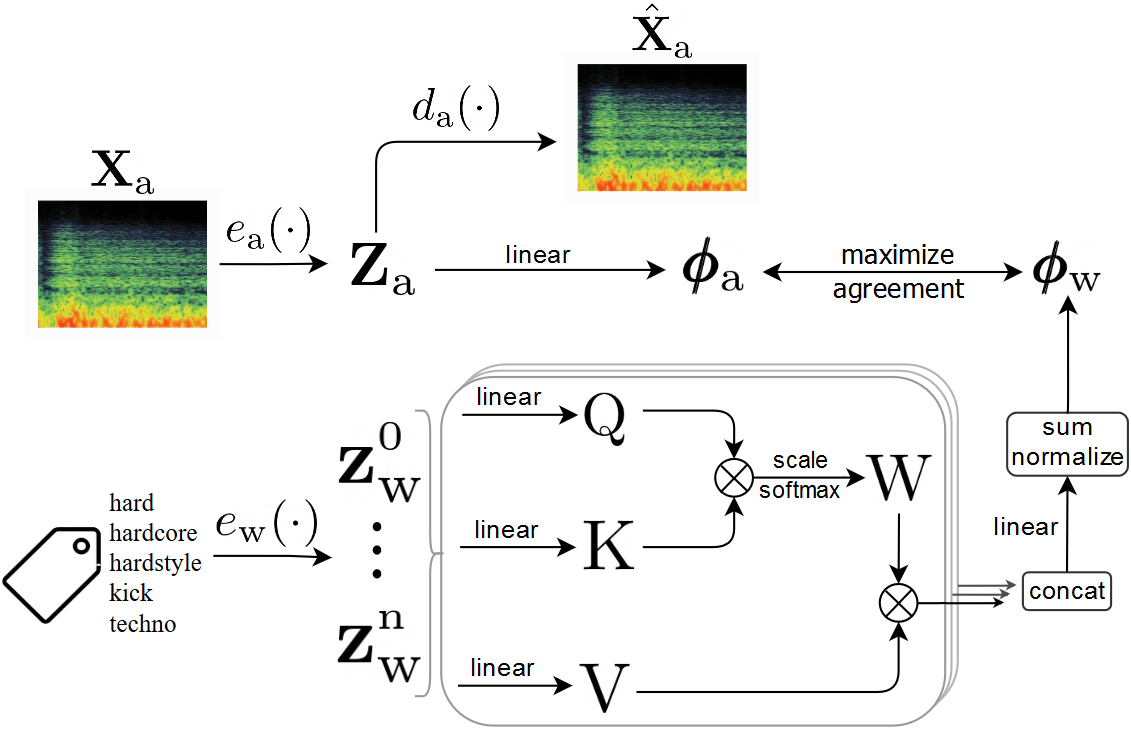}
  \caption{Illustration of our method. $\pmb{\phi}_{\text{a}}$ and $\pmb{\phi}_{\text{w}}$ are aligned by maximizing their agreement through contrastive learning and, at the same time, $\mathbf{Z}_{\text{a}}$ is used for reconstructing back the original spectrogram input. Word embeddings are passed through a multi-head scaled dot-product self-attention layer in order to build higher-level semantic vectors that are finally aggregated into a single vector $\pmb{\phi}_{\text{w}}$.}
  \label{fig}
\end{figure}
\vspace{-6pt}
\subsection{Multi-head, self-attention tags encoding}
\vspace{-3pt}
As our $e_{\text{w}}$ we select a pre-optimized word embedding model, using an embedding dimensionality of $F_{\text{w}}$, which outputs
$\mathbf{z}^{t_{\text{w}}}_{\text{w}}\in
\mathbb{R}^{F_{\text{w}}}$.
For processing the output of $e_{\text{w}}$ we follow the recent proposal of multi-head, self-attention in the Transformer model, where a scaled dot-product attention mechanism is employed to extract the relevant information of each word in a sentence~\cite{vaswani2017attention}. Since we use an unordered set of tags, we do not employ the positional embeddings for each tag. The multi-head self-attention attends on the set of encoded tags by the word embeddding model $e_{\text{w}}$, and extracts a contextual embedding of the set of tags. We use this contextual embedding as the latent representation of tags for our cross-modal alignment process with the contrastive loss.

Specifically, we employ three feed-forward neural networks, $\text{FNN}_{q}$, $\text{FNN}_{k}$, and $\text{FNN}_{v}$, $\text{FNN}_{o}$ (without non-linearities) as our linear transformations for the query, key, value and output of the self-attention mechanism, respectively. We follow the original paper of the self-attention~\cite{vaswani2017attention}, 
we use $H$ attention heads, and we apply the self-attention, Att, on the embeddings of tags, $\mathbf{Z}_{\text{w}}$. The output of Att is $\mathbf{O}\in\mathbb{R}^{H\times T_{\text{w}}\times F_{\text{w}}}$, according to~\cite{vaswani2017attention}. Then, we concatenate the result of each $H$, resulting to $\mathbf{O}'\in\mathbf{R}^{(H\cdot T_{\text{w}})\times F_{\text{w}}}$. Finally, to process the output of each attention head $H$ and obtain the contextual embedding for the input tags, $\pmb{\phi}_{\text{w}}$, we employ $\text{FNN}_{o}$ and a layer normalization process, $\text{LN}$, and we calculate the contextual embedding, $\pmb{\phi}_{\text{w}}\in\mathbb{R}^{V}$, as
\begin{align}
    \mathbf{O}' =& \text{FNN}_{o}(\mathbf{O})\text{ and}\\
    \pmb{\phi}_{\text{w}} =& \text{LN}(\sum\limits_{i=1}^{T_{\text{w}}}\mathbf{O}'_{i})\;\text{,}
\end{align}
\noindent
where $\mathbf{O}'\in\mathbb{R}^{T_{\text{w}}\times V}$ and $\text{FNN}_{o}$ has its weights shared along the $H\cdot T_{\text{w}}$ dimension. 

\subsection{Cross-modal alignment and optimization}
To align the learned latent representations from audio and tags, we employ a constrastive loss and we maximize the agreement of $\pmb{\phi}^{n_{\text{B}}}_{\text{a}}$ and $\pmb{\phi}_{\text{w}}^{n_{\text{B}}}$ on minibatches $\mathbb{G}_{\text{b}}$ of size
${N_{\text{b}}}$, randomly sampled from our dataset $\mathbb{G}$. 
To reduce the mismatch between the two different modalities, we employ a feed-forward neural network ($\text{FNN}_{\text{c-a}}$) to process $\mathbf{z}^{n_{\text{B}}}_{\text{a}}$, as
\begin{equation}
    \pmb{\phi}^{n_{\text{B}}}_{\text{a}} = \text{FNN}_{\text{c-a}}(\mathbf{z}^{n_{\text{B}}}_{\text{a}})\text{.}
\end{equation}
\noindent

Then, we employ a contrastive loss between $\pmb{\phi}^{n_{\text{B}}}_{\text{a}}$ and $\pmb{\phi}_{\text{w}}^{n_{\text{B}}}$ 
and we jointly optimize $e_{\text{a}}$, $d_{\text{a}}$, $\text{FNN}_{q}$, $\text{FNN}_{k}$, $\text{FNN}_{v}$, and $\text{FNN}_{o}$ by minimizing the loss
\begin{equation}
    \mathcal{L}(\mathbb{G}_{\text{b}}, e_{\text{a}}, d_{\text{a}}, \text{Att}) = \lambda_{\text{a}}\mathcal{L}_{\text{a}} +
    \lambda_{\xi}\mathcal{L}_{\xi}\text{,}
\end{equation}  
\noindent
where $\mathcal{L}_{\text{a}}$ is the generalized Kullback-Leibler divergence between $\mathbf{X}_{\text{a}}$ and $\hat{\mathbf{X}}_{\text{a}}$, shown to work good for audio reconstruction~\cite{drossos2018mad,mimilakis:2018:icassp}. $\mathcal{L}_{\xi}$ is the contrastive loss between paired $\pmb{\phi}^{n_{\text{B}}}_{\text{a}}$ and $\pmb{\phi}_{\text{w}}^{n_{\text{B}}}$ and other examples in the minibatch, as defined in~\cite{chen2020simple}. We use a temperature parameter $\tau$ for $\mathcal{L}_{\xi}$, and $\lambda_{\star}$ is a hyper-parameter used for numerical balancing of the two losses.

\vspace{-6pt}
\section{Evaluation}
\vspace{-6pt}
\label{sec:eval}
We evaluate our method by assessing the performance of $e_{\text{a}}$ as a pre-trained audio embedding extractor in different audio classification tasks.
%
We utilize a different audio dataset for each task and we compare the performance of our $e_{\text{a}}$ against COALA \cite{favory2020coala} and a set of hand-crafted MFCCs feature. We choose COALA because is the SOTA method that deals with cross-modal alignment of audio and text. 
\vspace{-6pt}
\subsection{Pre-training dataset and data pre-processing}
\vspace{-3pt}
%
%
We use the same dataset as used in COALA, which consists of sounds and associated tags collected from Freesound platform~\cite{font2013freesound}. 
We compute $F_{\text{a}}=96$ log-scaled mel-band energies using sliding windows of 1024 samples ($\approx$46 ms), with 50\% overlap and the Hamming windowing function.
Then, we select the spectrogram patch of size $T_{\text{a}}=96$ that has maximum energy in each sample.
This process leads to 189 896 spectrogram patches. 10\% of these patches are kept for validation and all the patches are scaled to values between 0 and 1.
%
We process the tags associated to the audio clips
by firstly removing any stop-words and making any plural forms of nouns to singular.
We remove tags that occur in more than 70\% of the sounds as they can be considered less informative, and consider the $C$=1000 remaining most occurring tags.
Then we train a continuous bag-of-words Word2Vec model~\cite{mikolov2013efficient} using these processed tags and we use this model as our $e_{\text{w}}$.
%

%
To have our method comparable with COALA and assess the impact of our proposed approach for learning contextual tags, we follow COALA and we use $N_{\text{e-a}}=5$ $\text{CNN}_{n_{\text{e-a}}}$, with $K_{\text{e-a}}=4$ and $S_{\text{e-a}}=2$. We use $C^{\text{in}}_{1} = 1$ and $C^{\text{out}}_{n_{\text{e-a}}}=128$. 
These hyper-parameters result to $V=1152$ and $e_{\text{a}}$ has approximately 2.4M parameters.
The tag encoder takes a set of $T_{\text{w}}=10$ maximum tags. It uses fully-connected linear transformations that retain the same dimension as the word embeddings, producing tag-based embedding vectors of the same dimension. We utilize two different $F_{\text{w}}$, one of 128 and another of 1152. The first of 128 is due to the fact that we are using a small scale vocabulary and we choose to have a small embedding size for $e_{\text{w}}$. The second, is for having $F_{\text{w}} = V$. Additionally, we employ two different set-ups for our Att, one with far and another with one attention head. We indicate the different combinations as ``w2v-$F_{\text{w}}$-Hh'', where H is the amount of attention heads. For example, ``w2v-128-1h'' means that we are using $F_{\text{w}}=128$ and one attention head. 
We also employ the self-attention strategy with a simple mean aggregation of the tags' word embeddings, which we refer as w2v-128-mean and w2v-1152-mean, to assess the impact of Att when using a simpler aggregation of its output. 
Finally, we employ the 20 first mel-frequency cepstral coefficients (MFCCs) with their $\Delta$s and $\Delta\Delta$s as a low anchor, using means and standard deviations through time, and we term this case as MFCCs.

All our models are trained for 200 epochs, using a minibatch size $N_{\text{B}}$=128 and an SGD optimizer with a learning rate value of 0.005.
We utilize the validation set to define the different $\lambda$'s at Eq.~(11)
and the constrastive loss temperature parameter $\tau$, to $\lambda_{\text{a}}$=5, $\lambda_{\xi}$=10, and $\tau=0.1$.
We use dropout probability of 0.25 for our $e_{\text{a}}$ and $d_{\text{a}}$ and 0.1 after the tag-based embedding model to avoid overfitting while training.
%
%

\vspace{-6pt}
\subsection{Audio-based classification}
\vspace{-6pt}
We assess the performance of the different embeddings extracted with $e_{\text{a}}$ in three different audio classification tasks. For each task, we employ the pre-trained, with our method, $e_{\text{a}}$, and we extract audio embeddings $\mathbf{z}_{\text{a}}$ for the the audio data of the corresponding task. Then, adopt the corresponding training protocol of the task (e.g. cross-fold validation) and we optimize a multi-layer perceptron (MLP) with one hidden layer of 256 features, similar to what is used in~\cite{cramer2019look, favory2020coala}. Finally, we assess the performance of the classifier using the data and the testing protocol of each task. To obtain an unbiased evaluation of our method, we repeat 10 times the training procedure of the MLP in each task and report the mean accuracy. Additionally, in order to understand if the self-attention mechanism is actually beneficial for obtaining a better semantic embedding from tags, we perform a tag-based classification for comparing the different approach versions.
For this purpose, we use the UrbanSound8K dataset and the tags of the associated samples from Freesound.

As the first task, we consider Sound Event Recognition (SER), where we use the UrbanSound8K dataset (US8K) \cite{salamon2014dataset}. US8K consists of around 8000 single-labeled sounds of maximum 4 seconds and 10 classes. We use the provided folds for cross-validation. We additonally consider the task of Music Genre Classification (MGC), where we use the fault-filtered version of the GTZAN dataset \cite{tzanetakis2002musical, kereliuk2015deep} consisting of music excepts of 30 seconds, single-labeled split in pre-computed sets of 443 songs for training and 290 for testing. Finally, we also consider the Musical Instrument Classification (MIC) task. We use the NSynth dataset \cite{engel2017neural} which consists of more than 300k sound samples organised in 10 instrument families. 
However, because we are interested to see how our models performs with relatively low amount of training data, we sample from NSynth a balanced set of 20k samples from the training set which correspond to approximately 7\% of the original set. The evaluation set is kept the same.

%
For the above tasks and datasets, we use non-overlapping audio frames that are calculated similarly to the pre-training dataset.
These frames are given as input to the different models in order to obtain audio embeddings.
In order to obtain fixed-length vectors, the audio embeddings are aggregated using a mean average statistic and finally used as an input to a classifier that is trained for each corresponding task.
Additionally, resulting embedding and MFCCs vectors are standardized to zero-mean and unit-variance, using statistics calculated from the training split of each task.
%
%

\vspace{-6pt}
\section{Results}
\vspace{-6pt}
\label{sec:results}
%
%
%
Table 1 shows the performances of the different embeddings our MFCCs baseline and previous results from COALA~\cite{favory2020coala}.
The self-attention mechanism used in the tag-based network benefits the classification performance in SER and MGC. 
This indicates that our proposed method indeed results in learning a contextual embedding that can be effectively used for learning better general audio representations. 
For MIC however, we do not observe any benefit from it.
A reason may be that the classification of a musical instrument does not rely much on the context of employed textual descriptions, at least for the musical instruments contained in the employed dataset for MIC. This means that that classifying instrument samples can be done by solely using representations learned by the audio autoencoder, without any semantic information. 
In \cite{favory2020coala}, it was observed that the reconstruction objective was bringing important improvements in this case, and probably, the enrichment of semantics achieved with the alignment with the tag-based latent representation losses its benefits.
%
Using more attention heads is able to bring better performance in SER and MGC. 
This suggests that the pre-trained word representation we use can benefit from more powerful attention mechanism.
%
The impact of the embedding size of the word embeddings is not clear from our experiment.
But, it shows that using different dimensions for the audio autoencoder and the tag-based encoder does not necessary hinders the contrastive alignment.
%

When using tags for performing the classification on US8K for SER, there is no benefit of using multiple attention heads, and the self-attention mechanism is only slightly improving the performance compared to the mean aggregation strategy.
Moreover, the results show that audio-based classification is still not performing as well as the tag-based one, which could suggest that more powerful audio encoders could be better aligned with the semantics of the content, and produce better results.

\begin{table}
\centering
\label{table:results}
\caption{Average mean accuracy for SER, MGC, and MIC. Additionally performances on US8K dataset using a tag-based classifier are reported in the last column. 
}
\begin{tabular}{l|ccc|c}
  & \textbf{SER} & \textbf{MGC} & \textbf{MIC} & \textbf{US8K-tag} \\
  \toprule
  MFCCs & 65.8 & 49.8 & 62.6 & - \\
  w2v-128-1h  & 71.5          & 61.3          & 68.9        & 79.2 \\
  w2v-1152-1h & 72.1          & 61.5          & 68.6        & \textbf{80.3} \\
  w2v-128-4h  & \textbf{73.5} & 59.6          & 69.7        & 79.4 \\
  w2v-1152-4h & 70.5          & \textbf{63.4} & 69.9        & 78.7 \\
  w2v-128-mean   & 71.3          & 60.4          & 70.0        & 79.7 \\
  w2v-1152-mean   & 71.1          & 60.7          & 68.4        & 78.5 \\
  COALA \cite{favory2020coala}    & 72.7          & 60.7        & \textbf{73.1}  & -
\vspace{-14pt}
\end{tabular}
\end{table}

\vspace{-6pt}
\section{Conclusion}
\vspace{-6pt}
\label{sec:conclusion}
In this work we present a method for cross-modal alignment of audio and tags.
The proposed approach uses a pre-trained word embedding and learns contextual tag embeddings that are aligned with audio embeddings using contrastive learning.
From audio samples and associated tags, our method is able to learn semantically enriched audio representation that can be used in different classification tasks.
The embedding model produced is evaluated in three different downstream tasks, including sound event recognition, music genre and music instrument classifications.
Over a previous similar method \cite{favory2020coala}, the proposed approach relies on pre-trained word embeddings that grants the the advantage of being directly able to be used in a wider range of applications, such as cross-modal retrieval or zero-shot learning.
Future work will focus on improving the model and evaluating in different scenarios, such as the one just mentioned.

\vfill

\bibliographystyle{IEEEbib}
\bibliography{refs}

\end{document}